\documentclass[letterpaper]{article}

\usepackage[T1]{fontenc}
\usepackage{geometry}
\usepackage{float}
\usepackage{amsmath}
\usepackage{amssymb}
\usepackage{graphicx}
\usepackage{tikz}
\usepackage[hidelinks]{hyperref}
\usetikzlibrary{calc,positioning,arrows.meta,decorations.pathreplacing}

\usepackage{achemso}
\setkeys{acs}{articletitle=true}

\usepackage{authblk}
\title{Physics-constrained neural networks for surrogate modeling of lossless periodic structures}
\author[1]{Eric Prehn*}
\author[1]{Peter Jung}
\affil[1]{Institute of Space Research, German Aerospace Center (DLR), Berlin, Germany}
\date{*Email: eric.prehn@dlr.de}

\begin{document}
\maketitle

\begin{abstract}
We introduce a physics-constrained neural network for the rapid prediction of rigorous coupled-wave analysis outputs in the form of Jones matrices. Starting from energy conservation in lossless layered periodic structures, we use the fact that the scattering outputs lie on a Stiefel manifold. This energy constraint is enforced as a hard condition by projecting onto the manifold using differentiable symmetric orthogonalization. The resulting surrogate enforces energy conservation by construction while preserving differentiability for gradient-based inverse design. The performance and generality of the proposed approach are demonstrated through the inverse design of a diffractive waveguide combiner for augmented reality glasses.
\end{abstract}

\section*{Keywords}
physics-constrained neural networks; rigorous coupled-wave analysis; surrogate modeling; inverse design; diffractive optics; augmented reality

\section{Introduction}
Photonics engineering leverages the geometric design of materials to control electromagnetic (EM) responses. Periodic structures are widely employed in photonic systems, with applications such as beam splitters \cite{Langois90_multiplebeams}, beam steering \cite{Yang2018_BeamSteering,Zhang2018_HygensMetaoptics}, guided-mode resonance filters \cite{Wang93_guidedmoderesfilters}, couplers in integrated optics \cite{Taillaert2006_gratingcouplers,Roelkins2007_High}, imaging \cite{Lee2002_Blazedimaging} and waveguide combiners \cite{Kress2020_WaveguideCombiners}. The design of these structures relies on accurate and efficient EM simulations, which are essential for both forward modeling and inverse design workflows.

Rigorous coupled-wave analysis (RCWA) \cite{Moharam1981_RCWA,Moharam1995_Formulation,Moharam1995_Stable} provides a semi-analytic solution for layered periodic structures and is widely used in photonic device design. However, its computational cost can become prohibitive for large-scale optimization and inverse design over high-dimensional design spaces. Neural networks (NNs) offer a promising alternative as surrogate EM solvers, learning from data to accelerate simulations by orders of magnitude \cite{Nadel2019_Deep,Liu2018_Training}. Once trained, they enable rapid inference and flexible inverse design through readily accessible gradients \cite{Tahersima2019_DeepNNIC,Peurifoy2018_NanophotonicANNs,An2019_DLapproach_alldielectric}.

To enforce consistency with the underlying physical laws, several physics-informed and physics-augmented neural-network approaches have been proposed for EM modeling \cite{Chen2020_PINNnanoOptics,Lim2022_MaxwellNet}. Physics-informed neural networks typically incorporate governing equations through residual terms in the loss function \cite{Raissi2019_Physics}, while physics-augmented surrogate models embed domain knowledge through tailored architectures or hybrid solvers such as WavYNet \cite{Chen2022_WaveYNet}. These methods improve physical fidelity but often rely on soft constraints and weighted loss terms.

Here, we introduce a physics-constrained neural network (PCNN) that enforces energy conservation as a hard output constraint. For lossless periodic structures, energy conservation restricts Jones matrices describing the amplitude, phase, and polarization of  scattered far fields to a Stiefel manifold \cite{Edelmann1998_Stiefel}. By projecting the network output directly onto the manifold, energy conservation is satisfied by construction. We evaluate the resulting surrogate against an unconstrained NN baseline to assess the impact of the energy constraint, and demonstrate its scalability in a large-scale waveguide-combiner optimization. The computational efficiency of the PCNN enables practical optimization at this scale and highlights the potential of the framework for designing augmented reality (AR) display systems. More broadly, it establishes a general methodology for the surrogate modeling and gradient-based design of lossless layered periodic structures.

\section{Energy-Conserving Jones Matrix Representation}
For a lossless periodic structure, the sum of diffraction efficiencies over all $M$ propagating reflected and transmitted orders satisfies
\begin{equation}\label{eq:DE}
\sum_{m=1}^M \eta_m = 1.
\end{equation}
Here, $\eta_m$ is the diffraction efficiency of the non-evanescent order $m$, defined as the ratio between the flux of the diffracted Poynting vector and the flux of the incident Poynting vector \cite{Hugonin2021_reticolo}. In direct RCWA calculations, this condition must be satisfied, with Moharam \emph{et al.} recommending a numerical accuracy on the order of $10^{-10}$ \cite{Moharam1995_Formulation}. Strict energy conservation is essential, as even a single RCWA prediction in which the total diffraction efficiency exceeds unity can compromise an entire optical simulation.

For an incident electric field with Jones vector $e$, the field diffracted into the $m$th order is
\begin{equation}
e_m = J_m 
e =
\begin{pmatrix}
j^{\mathrm{TE},\mathrm{TE}}_m & j^{\mathrm{TM},\mathrm{TE}}_m \\
j^{\mathrm{TE},\mathrm{TM}}_m & j^{\mathrm{TM},\mathrm{TM}}_m
\end{pmatrix}
\begin{pmatrix}
e_{\mathrm{TE}} \\
e_{\mathrm{TM}} 
\end{pmatrix} =
\begin{pmatrix}
a_m & b_m \\
c_m & d_m
\end{pmatrix}e.
\end{equation}
The Jones matrix $J_m \in \mathbb{C}^{2 \times 2}$ is expressed in transverse electric (TE) and transverse magnetic (TM) polarization bases, defined with respect to planes of incidence and diffraction~\cite{Hugonin2021_reticolo}. For clarity, we adopt the shorthand notation $a_m, b_m, c_m, d_m$ for matrix elements. The condition in Eq.~(\ref{eq:DE}) is equivalent to
\begin{equation}
\sum_{m=1}^M \|J_m e\|^2
= e^\dagger\left(\sum_{m=1}^{M} J_m^\dagger J_m\right)e
\overset{\displaystyle !}{=} \|e\|^2,
\end{equation}
which must hold for arbitrary $e$, and hence
\begin{equation}\label{eq:constraints}
\sum_{m=1}^{M} J_m^\dagger J_m =
\begin{pmatrix}
\sum_{m=1}^{M} |a_m|^2 + |c_m|^2 & \sum_{m=1}^{M} a_m^* b_m + c_m^* d_m \\
\sum_{m=1}^{M} b_m^* a_m + d_m^* c_m & \sum_{m=1}^{M} |b_m|^2 + |d_m|^2
\end{pmatrix}
=
\begin{pmatrix}
1 & 0 \\
0 & 1
\end{pmatrix}=\mathbb{I}_2.
\end{equation}
A convenient construction satisfying these constraints is obtained by forming vectors of Jones matrix elements,
\begin{equation}\label{eq:uv_def}
\begin{aligned}
u &= (a_1,a_2,\ldots,a_M,c_1,c_2,\ldots,c_M), \\
v &= (b_1,b_2,\ldots,b_M,d_1,d_2,\ldots,d_M).
\end{aligned}
\end{equation}
Enforcing orthonormality of $u$ and $v$ ensures that Eqs.~(\ref{eq:DE}) and~(\ref{eq:constraints}) are satisfied:
\begin{equation}\label{eq:constraints-uv}
\sum_{m=1}^{M} J_m^\dagger J_m =
\begin{pmatrix}
\|u\|^2 & u^\dagger v \\
 v^\dagger u & \|v\|^2
\end{pmatrix}
=
\begin{pmatrix}
1 & 0 \\
0 & 1
\end{pmatrix}.
\end{equation}

\section{Physics-Constrained Neural Network}
The surrogate model aims to learn the nonlinear mapping from RCWA input parameters to the corresponding Jones matrices,
\begin{equation}\label{eq:mapping}
(p_1,p_2,\ldots,p_D,k_x,k_y) \mapsto \{J_m(p_1,p_2,\ldots,p_D,k_x,k_y)\}_{m=1}^M.
\end{equation}
Here, $p_d$ denotes a geometric design parameter, such as grating duty cycle, $D$ is the dimensionality of the design space, and $(k_x,k_y)$ are the transverse wavevector components of the incident light. Each input configuration is mapped to Jones matrices for $M$ diffraction orders retained from the RCWA simulation. The layered periodic structure is invariant along the longitudinal $z$-direction within each layer, and periodic in the transverse plane.  We propose a PCNN architecture that first predicts an unconstrained complex matrix $V=[u\;v]$ and then projects it onto the Stiefel manifold $\mathrm{St}(2,2M,\mathbb{C})$, the set of orthonormal 2-frames in $\mathbb{C}^{2M}$ \cite{Edelmann1998_Stiefel}. At the output of the PCNN, the resulting orthonormal vectors $\tilde{V}=[\tilde{u}\;\tilde{v}]$ are reshaped into Jones matrices.

The projection is performed using Löwdin symmetric orthogonalization \cite{Löwdin1970_ortho3,Aiken1980_ortho}. Writing the thin singular-value decomposition (SVD) \cite{Courant1989_Methods} of $V$  as
\begin{equation}
V = U\Sigma W^\dagger,
\end{equation}
the orthonormalized output is
\begin{equation}\label{eq:NN_svd_out}
\tilde{V} = U W^\dagger.
\end{equation}
This projection treats the two output vectors symmetrically, in contrast to sequential Gram--Schmidt orthogonalization \cite{Courant1989_Methods}. For further details, a derivation of Eq.~(\ref{eq:NN_svd_out}) is provided (Section S1 of the Supporting Information). In summary, the network learns within the manifold of admissible solutions, guaranteeing that the final outputs satisfy energy conservation. 

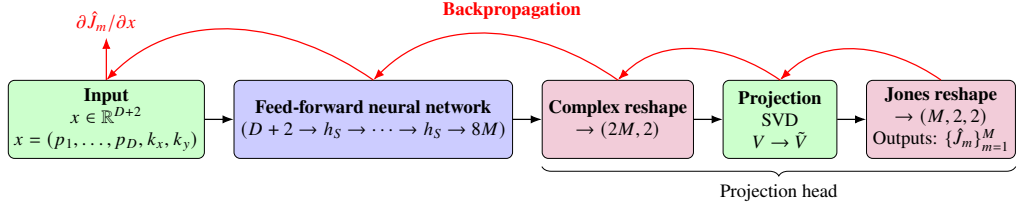
\begin{figure}[t]
\centering
\resizebox{\linewidth}{!}{%
\begin{tikzpicture}[
    font=\large,
    box/.style={draw,rounded corners,minimum height=1.75cm,
        minimum width=2.55cm,inner sep=4pt,align=center},
    arrow/.style={-{Latex[length=2.5mm]}, thick},
    backarrow/.style={-{Latex[length=2.5mm]}, thick, red},
    node distance=0.65cm
]
\node[box, fill=green!20] (input) {\textbf{Input}\\
$x \in \mathbb{R}^{D+2}$\\
$x=(p_1,\ldots,p_D,k_x,k_y)$};

\node[box, fill=blue!20, right=of input, minimum width=5.2cm] (mlp) {
\textbf{Feed-forward neural network}\\
$(D+2 \rightarrow h_S \rightarrow \cdots \rightarrow h_S \rightarrow 8M)$};

\node[box, fill=purple!20, right=of mlp] (reshape1) {
\textbf{Complex reshape}\\
$\rightarrow (2M,2)$};

\node[box, fill=green!20, right=of reshape1] (projection) {
\textbf{Projection}\\
SVD\\
$V\rightarrow\tilde{V}$};

\node[box, fill=purple!20, right=of projection] (reshape2) {
\textbf{Jones reshape}\\
$\rightarrow (M,2,2)$\\
Outputs: $\{\hat{J}_m\}_{m=1}^M$};

\node[red] at ($(input.west)!0.5!(reshape2.east) + (0,2.45cm)$)
{\large\textbf{Backpropagation}};

\draw[arrow] (input) -- (mlp);
\draw[arrow] (mlp) -- (reshape1);
\draw[arrow] (reshape1) -- (projection);
\draw[arrow] (projection) -- (reshape2);

\draw[backarrow] (input.north) -- ++(0,0.95cm)
node[above] {\large $\partial \hat{J}_m/\partial x$};
\draw[backarrow] (reshape2.north) to[out=145,in=45] (projection.north);
\draw[backarrow] (projection.north) to[out=145,in=45] (reshape1.north);
\draw[backarrow] (reshape1.north) to[out=145,in=45] (mlp.north);
\draw[backarrow] (mlp.north) to[out=145,in=45] (input.north);

\draw[decorate,decoration={brace,mirror,amplitude=6pt}]
([yshift=-5pt]reshape1.south west) -- ([yshift=-5pt]reshape2.south east)
node[midway, below=8pt, font=\large] {Projection head};
\end{tikzpicture}%
}
\caption{PCNN architecture with projection onto the Stiefel manifold. Red arrows denote gradient backpropagation, while the projection head enforces energy conservation by construction and preserves end-to-end differentiability.}
\label{fig:nn-architecture}
\end{figure}

In this work, the models are implemented as fully connected multilayer perceptrons (Figure~\ref{fig:nn-architecture}) with rectified linear unit activations \cite{Agarap2019_Relu}. The projection is independent of the particular network architecture or input parameterization and provides a general approach for incorporating energy conservation into RCWA surrogate models.

\section{Dataset Generation and Model Training}
\begin{figure}[t]
\centering
\includegraphics[width=\linewidth]{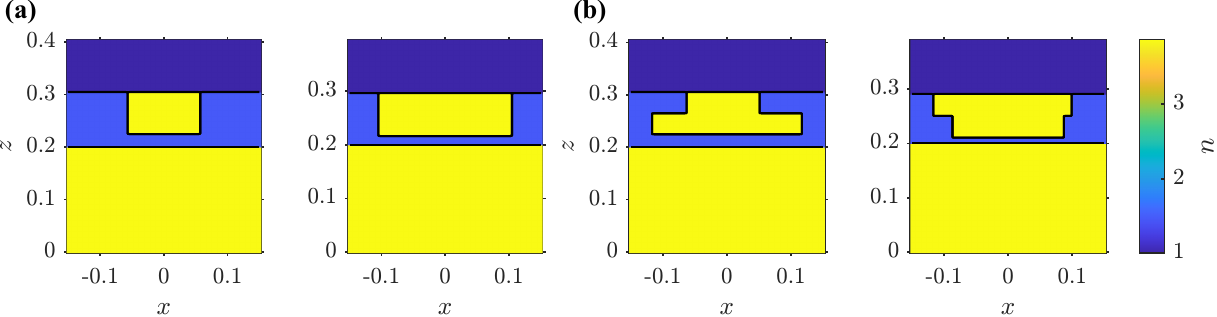}
\caption{Representative all-dielectric grating structures and their spatial refractive-index distributions $n(x,z)$ at $\lambda=0.63\,\mu\mathrm{m}$ from the (a) $D=2$ and (b) $D=4$ design spaces.}
\label{fig:grats_datasets}
\end{figure}

In this work, we investigate the surrogate modeling of one-dimensional all-dielectric binary gratings composed of single-crystal silicon and silicon dioxide. These materials were modeled as lossless owing to their low optical loss across the visible regime \cite{Sell2016_visible,Gao2013_Si02} and were selected for their compatibility with standard semiconductor fabrication processes. RCWA simulations were performed using Reticolo \cite{Hugonin2021_reticolo} to generate datasets for the two design spaces. The first design space, denoted $D=2$, comprises a grating parameterized by the duty cycle ($DC$) and the thickness of a $\mathrm{SiO}_2$ underlayer ($h_{\mathrm{ul}}$). The $D=4$ design extends the structure to a multilevel configuration, in which an additional binary grating is introduced on top of the first. Its geometric parameters include the duty cycles of the two gratings, $h_{\mathrm{ul}}$ and the lateral displacement of the upper grating. Datasets comprising 25{,}000 and 100{,}000 samples were generated for the $D=2$ and $D=4$ design spaces, respectively, with samples spanning a range of incident wavevectors (Section S2 of the Supporting Information). Representative gratings from the datasets are shown in Figure~\ref{fig:grats_datasets}, and these datasets are subsequently used in the waveguide optimization demonstration.

The models are trained using supervised learning (training details are provided in Section S3 of the Supporting Information), and performance is evaluated by comparing the predicted Jones matrices $\hat{J}_m$ to the simulated RCWA matrices $J_m$ using

\begin{equation}
\label{eq:NN_loss}
\mathrm{MSE} =
\frac{1}{4NM}
\sum_{n=1}^{N}
\sum_{m=1}^{M}
\sum_{i,j \in \{\mathrm{TE},\mathrm{TM}\}}
\left| 
\hat{J}_m^{n,ij}-J_m^{n,ij}
\right|^2
.
\end{equation}
The mean-squared error (MSE) captures deviations in both amplitude and phase via the real and imaginary components across all $N$ samples and $M$ diffraction orders. Unlike related approaches that rely on multiple weighted loss terms with potentially competing gradients~\cite{Chen2022_WaveYNet,Prehn2025_Accelerating}, the PCNN employs a single loss term. This avoids gradient-balancing issues and eliminates the need to tune loss-term weights.

To test the impact of the constraint, the PCNN models were compared with unconstrained NN models. Both were trained with the same supervised MSE objective and differ only by the hard projection layer. The NN uses the same input features and hidden-layer form, but omits the projection step and directly outputs Jones matrices.
To ensure a fair comparison, the NN and PCNN were independently selected using the same validation-based model-selection protocol. For both models, the number of hidden layers $S$ and number of nodes $h_S$ in each layer were swept over an identical search space, and this procedure was repeated for different fractions of the training dataset (Section S3 of the Supporting Information). The models with the lowest validation MSE were selected and evaluated on held-out test sets.

\section{Impact of the Energy Constraint}
Here, we evaluate the models trained on the complete training dataset and report on reduced training-data regimes in Section S3 of the Supporting Information. Across all three wavelengths and both design spaces, the PCNN achieved a lower test MSE than the NN, indicating that the energy-conservation constraint acts as an inductive bias. Although both methods achieve similar test MSE values on the order of $10^{-5}$ to $10^{-6}$ (Section S3 of the Supporting Information), substantial differences in physical consistency remain. To quantify this, the energy conservation for a structure is measured by the deviation of the predicted response from Eq.~(\ref{eq:constraints}),
\begin{equation}\label{eq:energy_error}
    \Delta_E = 
\left\| 
\sum_{m=1}^{M} \hat{J}_m^{\dagger} \hat{J}_m - \mathbb{I}_2
\right\|_{\infty} = \;  \max_{i,j} \left| \left( \sum_{m=1}^{M} \hat{J}_m^{\dagger} \hat{J}_m - \mathbb{I}_2\right)_{ij} \right| .
\end{equation}
Here, $\Delta_E=0$ corresponds to exact conservation of total diffracted energy for arbitrary incident polarization. Because the PCNN output is projected onto the Stiefel manifold, Eq.~(\ref{eq:energy_error}) is satisfied up to numerical precision, resulting in test-set energy-conservation errors of order $\Delta_E\sim10^{-6}$.  In comparison, the unconstrained NN yields errors of order $\Delta_E\sim10^{-2}$. This discrepancy highlights the fact that low prediction error does not necessarily imply physical consistency. For lossless structures, even small deviations from the admissible manifold can produce significant violations of energy conservation, limiting the reliability of unconstrained surrogate models.

\begin{figure}[H]
\centering
\includegraphics[width=\linewidth]{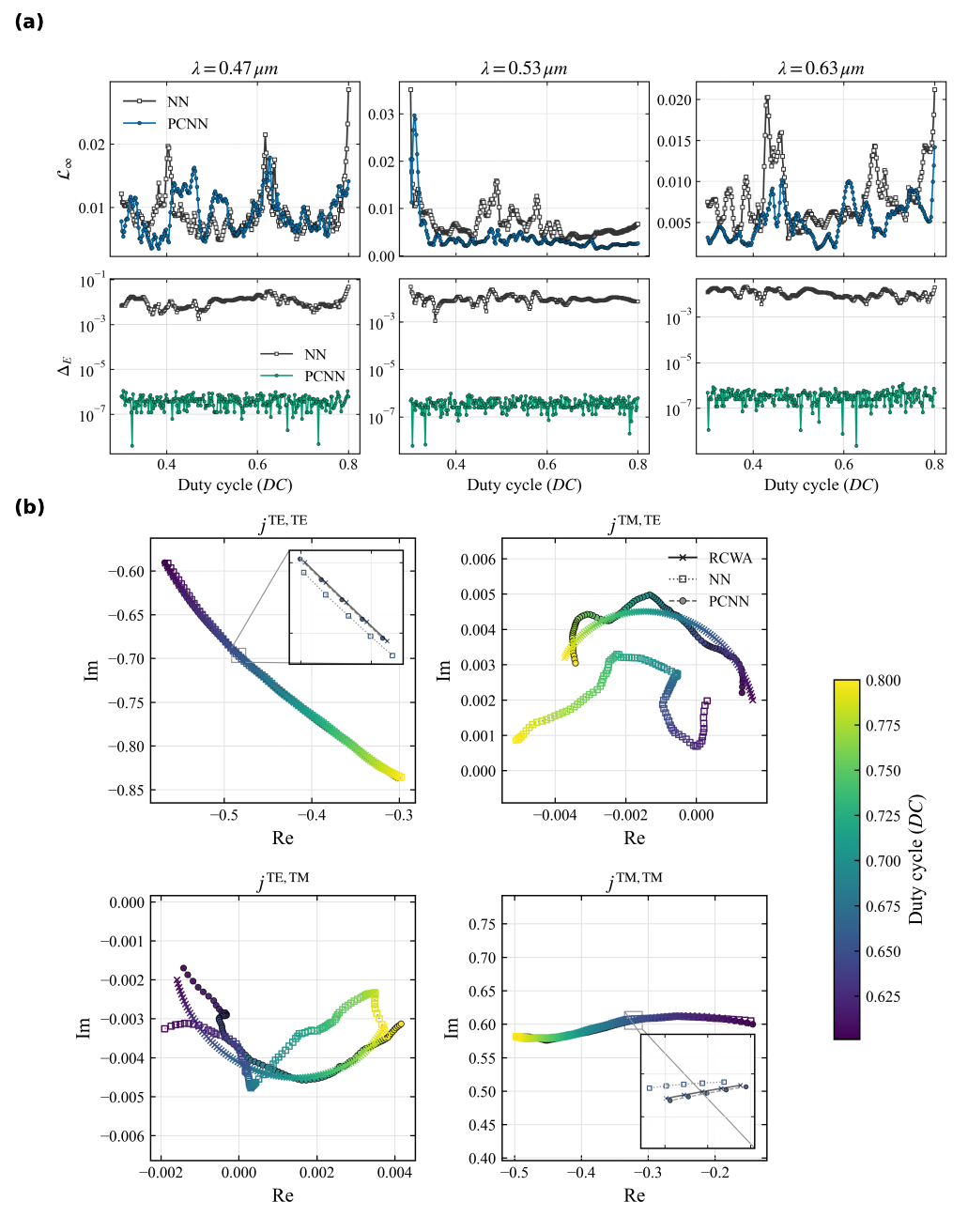}
\caption{Duty-cycle test set sweep for the $D=2$ design space at fixed underlayer thickness $h_{\mathrm{ul}}=0.025~\mu\mathrm{m}$ under fixed illumination. 
(a) $\mathcal{L}_\infty$ error and energy-conservation violation for NN and PCNN models across three wavelengths.
(b) Complex-plane Jones-matrix trajectories of the zeroth reflected order for $\lambda=0.53\,\mu\mathrm{m}$, comparing simulated RCWA with surrogate predictions.}
\label{fig:jones_dc}
\end{figure}
  
The accuracy of the surrogate model depends on several factors, including model expressivity \cite{yarotsky2017_errorbounds}, sampling density, and the underlying physics within the design space. The sensitivity of the Jones matrices obtained from converged RCWA solutions can vary considerably across the parameter space $(p_1,\ldots,p_D,k_x,k_y)$. Such sensitivity can arise, for example, near guided-mode resonances \cite{Wang90_guided_mode_reso} and diffraction-order cutoffs \cite{Hessel65_woods}, where small geometric or angular variations can lead to large changes in the optical response. For the grating structures considered here, the duty cycle is a sensitive parameter as it directly modifies the Fourier coefficients of the periodic permittivity distribution in the RCWA formulation \cite{Moharam1981_RCWA,Moharam1995_Formulation,Moharam1995_Stable}, reshaping the Fourier spectrum of the layer. We therefore perform one-dimensional duty-cycle sweeps on additional test sets while keeping the remaining geometric and wavevector parameters fixed. With reliability in mind and to assess worst-case behavior beyond MSE, we evaluate the infinity norm error for a structure as
\begin{equation} \label{eq:J_infty_error}
\mathcal{L}_{\infty} = 
\max_{m,i,j}
\left|
\hat{J}^{ij}_m - J^{ij}_m
\right|.
\end{equation}

The $\mathcal{L}_\infty$ error corresponds to the maximum absolute element-wise error across all Jones matrix elements and diffraction orders. Figure~\ref{fig:jones_dc} shows the corresponding behavior for a sweep in the $D=2$ space. The upper panels compare the $\mathcal{L}_\infty$  and energy conservation errors as a function of duty cycle. Although the PCNN and NN achieve comparable Jones-matrix errors, the NN exhibits substantial violations of energy conservation, resulting in nonphysical loss and gain. The lower panels show the evolution of the zeroth reflected-order Jones matrix as a function of duty cycle for $\lambda=0.53~\mu\mathrm{m}$. Both models accurately reproduce the RCWA trajectory, while the PCNN predictions remain restricted to the manifold. 

Data-efficiency comparisons further show that the constraint usually preserves or improves predictive accuracy once the design spaces in this work are sufficiently sampled (Section S3 of the Supporting Information). Overall, the MSE performance and results along selected sensitive trajectories support the accuracy and physical reliability of the PCNNs. We next demonstrate their use in an inverse-design application.

\section{Waveguide Optimization Demonstration}

\begin{figure}[H]
\centering
\includegraphics[width=3.33in]{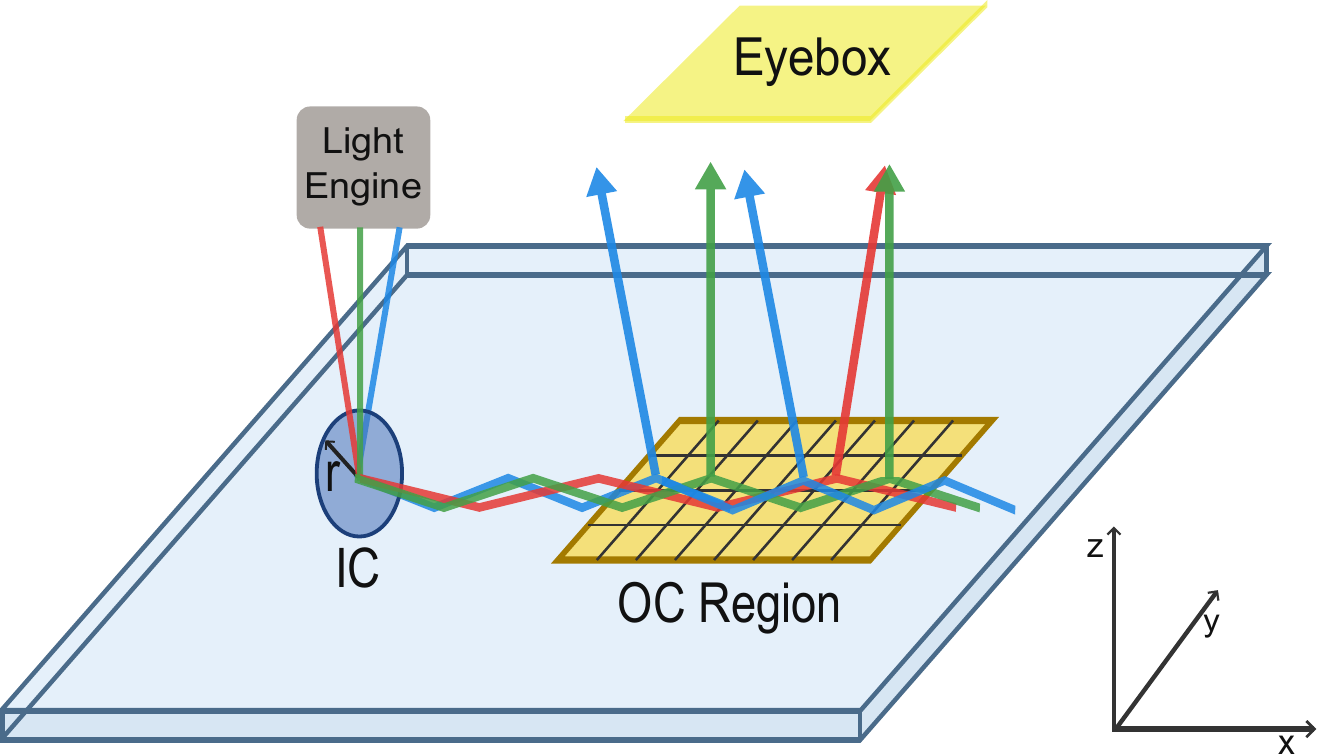}
\caption{Schematic diagram of the diffractive waveguide combiner. The incoupling (IC) region is a circle of radius $r$ with a fixed IC grating. The outcoupling (OC) region contains a grid of OC subregions, within each of which the grating shape is optimized.}
\label{fig:waveguide_diagram}
\end{figure}

To illustrate the capabilities of the proposed PCNN framework, we demonstrate a ray-tracing-based optimization of the outcoupling (OC) region in a full-color diffractive waveguide combiner for AR glasses. Gratings are located on the upper surface of a planar single-crystal silicon waveguide and form an incoupling (IC) region and an OC region, resulting in one-dimensional exit-pupil expansion \cite{Ding2023_ARreview,Kress2020_WaveguideCombiners}. The IC grating design is fixed, whereas the OC region is discretized into a two-dimensional grid of $40{,}000$ grating subregions. The OC region is modeled using a locally periodic approximation (LPA) \cite{Pestourie2018_Inverse}, assuming slowly varying grating geometries across neighboring subregions. Each OC subregion is assigned a periodic grating with parameter vector $\mathbf{p}\in\mathbb{R}^D$ and the collection of parameter vectors across all subregions defines a large design space $P$. To mimic manufacturable designs, selected geometric parameters are constrained to remain uniform across the OC region. Figure~\ref{fig:waveguide_diagram} sketches the waveguide layout and eyebox region, where rays are collected to form a viewable image.

The forward model combines geometric ray tracing with PCNN evaluations in a reduced-order treatment of the dominant waveguide propagation paths (Section S5 of the Supporting Information). It starts by tracing rays originating from a microdisplay, spanning a $20^\circ$ diagonal field of view (FOV). These rays enter the IC and undergo total internal reflection within the waveguide. When rays reach the discretized OC region, ray--grating interactions are evaluated using PCNNs. Relevant Jones matrices are applied sequentially to accumulate the transmitted and reflected fields, yielding intensity distributions across the eyebox. The polarization-averaged intensities are binned by wavevector direction and form the basis of the optimization objective, denoted $C(P)$. The objective uses approximate CIE 1931 color-matching functions~\cite{Wyman2013_Simple} to convert the simulated three-wavelength intensities into XYZ tristimulus values from which CIE-style $L^*a^*b^*$ chromatic coordinates are computed relative to a D65 white point. Nonzero $a^*$ and $b^*$ values are penalized, together with luminance and chromaticity variations across angular eyebox bins (Section S5 of the Supporting Information).

In conventional EM solvers, evaluating the optical response for each ray--grating interaction within an iterative loop can incur substantial computational cost. Here, the PCNNs enable efficient simulation of large ray ensembles while preserving energy conservation (time and memory complexity comparisons are provided in Section S4 of the Supporting Information). Forward and gradient computations are implemented within the JAX framework \cite{Bradbury2018_Jax}, enabling reverse-mode (adjoint) automatic differentiation, just-in-time compilation, and efficient vectorization over rays. The purpose of the demonstration is to illustrate the scale of differentiable optimization enabled by the surrogate framework rather than to provide a production-ready waveguide design. The optimization launches 40,015 rays per wavelength into the waveguide. Each iteration evaluates approximately $5\times10^6$ Jones matrices, recursively propagates electric fields, computes eyebox intensity distributions, evaluates adjoint gradients, and updates the design parameters. When JIT-compiled on an A100 GPU, the complete iteration requires approximately $0.5$ seconds. This computational performance enables repeated optimization studies under varying initial conditions and objective functions $C(P)$.

\begin{figure}[b!]
\centering
\includegraphics[width=\linewidth]{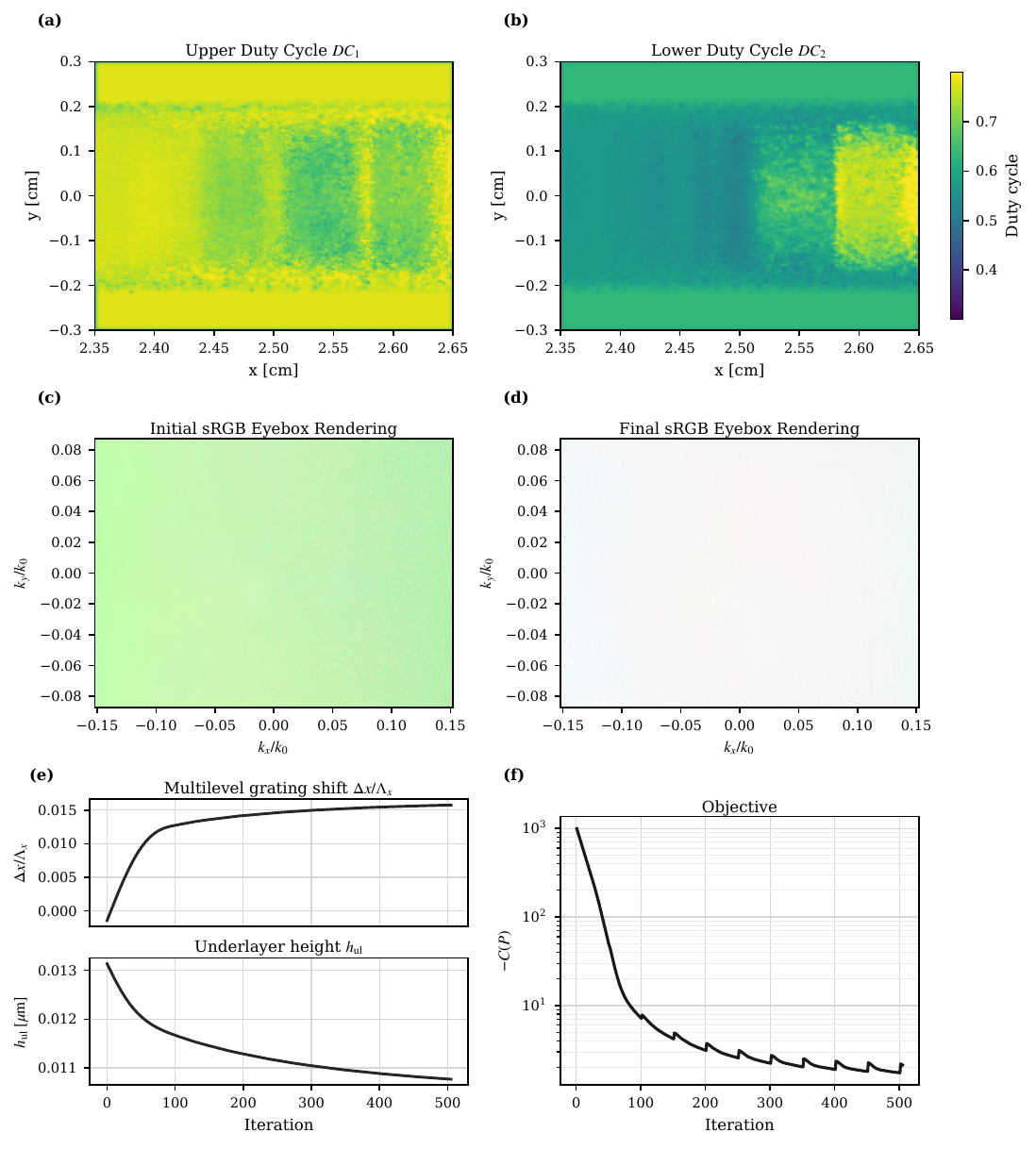}
\caption{Waveguide optimization in $D=4 $ design space. (a,b) Final spatial distributions of the upper and lower grating duty cycles, across the OC region. (c,d) Initial and final polarization-averaged eyebox renderings, obtained by converting the simulated RGB wavelength-channel intensities to sRGB for visualization. (e) Optimization trajectories of the spatially constant design parameters. (f) Objective trajectory over iterations.}
\label{fig:optimization_result}
\end{figure}

Figure~\ref{fig:optimization_result} shows the results of an optimization for the $D=4$ design space. The final lower and upper grating duty-cycle maps exhibit spatially varying structures across the OC region, reflecting the field redistribution required to homogenize the eyebox. Polarization-averaged eyebox renderings are shown in display sRGB for visualization purposes (Section S5 of the Supporting Information). The optimized design produces a more uniform full-color eyebox, which is consistent with the optimization objective. Video S1 provides an animation of the optimization in Figure~\ref{fig:optimization_result}, along with additional optimization runs. The $D=4$ design space yielded more uniform eyebox distributions than the $D=2$ design space while maintaining a comparable optimization runtime.

\section{Conclusion}

We presented a physics-constrained neural network for the surrogate modeling of RCWA outputs in the form of Jones matrices. Our results demonstrate that an unconstrained surrogate can exhibit substantial violations of energy conservation despite achieving low prediction error. In contrast, the Stiefel-manifold projection enforces energy conservation without compromising predictive accuracy while retaining full differentiability. This is relevant for optimization, where designs that violate energy conservation may otherwise be favored by the objective function. The PCNN therefore provides a more reliable surrogate for the simulation and optimization of lossless periodic structures.

Representing the optical response by Jones matrices is well suited to applications where only the far-field scattering response is required, retaining amplitude, phase, and polarization information without requiring the full electromagnetic field distribution. We demonstrated the computational scalability of this approach through the optimization of a full-color diffractive waveguide combiner involving millions of Jones-matrix evaluations per iteration.

Although RCWA was used here to generate the training data, the method may also be applied to other electromagnetic solvers, provided that the response can be expressed in terms of energy-conserving Jones matrices. The framework is therefore not limited to the structures considered here and provides an approach for the simulation and design of a broad range of lossless periodic structures.

\section*{Data Availability}
The RCWA-generated datasets used for training and evaluating the surrogate models are not publicly available at this time but may be obtained from the authors upon reasonable request.

\section*{Code Availability}
The surrogate-model implementations are not publicly available at this time but may be obtained from the authors upon reasonable request.

\section*{Supporting Information}
See the Supporting Information document (PDF) for symmetric orthogonalization; RCWA dataset generation; neural-network model selection and data-efficiency results; computational-complexity analysis; and details of the waveguide optimization. See Video S1 (MP4) for animation of the waveguide optimization. Both are included as ancillary files with this arXiv submission.

\section*{Funding}
This project was made possible by the DLR Quantum Computing Initiative and the Federal Ministry for Research, Technology and Space.

\newpage

\bibliography{PCNN_final}

\end{document}